\documentclass[english,submission,copyright,creativecommons]{eptcs}
\usepackage[T1]{fontenc}
\usepackage[latin9]{inputenc}
\usepackage{array}
\usepackage{multirow}
\usepackage{graphicx}
\usepackage{eurosym}
\usepackage{threeparttable}



\usepackage{babel}

\author{Gustavo Gonzalez-Granadillo
\institute{Institut Mines-T\'{e}l\'{e}com, T\'{e}l\'{e}com SudParis,
  CNRS UMR 5157 SAMOVAR\\  9 rue Charles Fourier, 91011 Evry, France}
\email{gustavo.gonzalez\_granadillo@telecom-sudparis.eu}
\and
Christophe Ponchel
\institute{Cassidian CyberSecurity\\
1 bd Jean Moulin, 78996 Elancourt Cedex, France}
\email{christophe.ponchel@cassidian.com}
\and
Gregory Blanc
\institute{Institut Mines-T\'{e}l\'{e}com, T\'{e}l\'{e}com SudParis,
  CNRS UMR 5157 SAMOVAR\\  9 rue Charles Fourier, 91011 Evry, France}
\email{gregory.blanc@telecom-sudparis.eu}
\and
Herv\'{e} Debar
\institute{Institut Mines-T\'{e}l\'{e}com, T\'{e}l\'{e}com SudParis,
  CNRS UMR 5157 SAMOVAR\\  9 rue Charles Fourier, 91011 Evry, France}
\email{herve.debar@telecom-sudparis.eu}
}

\def\titlerunning{Combining Technical and Financial Impacts for Countermeasure Selection}
\def\authorrunning{G. Gonzalez-Granadillo, C. Ponchel, G. Blanc, H. Debar}

\begin{document}

\title{Combining Technical and Financial Impacts for Countermeasure Selection}
\maketitle
\begin{abstract}
Research in information security has generally focused on providing a
comprehensive interpretation of threats, vulnerabilities, and attacks,
in particular to evaluate their danger and prioritize responses
accordingly. Most of the current approaches propose advanced
techniques to detect intrusions and complex attacks but few of these
approaches propose well defined methodologies to react against a given
attack. In this paper, we propose a novel and systematic method to
select security countermeasures from a pool of candidates, by ranking
them based on the technical and financial impact associated to each
alternative. The method includes industrial evaluation and simulations
of the impact associated to a given security measure which allows to
compute the return on response investment for different candidates. A
simple case study is proposed at the end of the paper to show the
applicability of the model.
\end{abstract}

\section{Introduction}
Innovation in Information Technology has brought numerous advancements but also some consequences. Cyber attacks have evolved along with technology, reaching a state of high efficiency and performance. Current research focuses on approaches to detect such attacks and demonstrate their strengths and the difficulty to mitigate them \cite{Agarwal10, Vetillard10}. Most of these works propose approaches to detect complex attacks but few of them propose a methodology to react against them.

In addition, research in dynamic response proposes automatic response mechanisms (e.g., the adaptation of security policies) to overcome the limitations of manual responses. However, these approaches remain limited since they do not analyse the impact of the selected countermeasures \cite{Deb07}. Inappropriate selection of countermeasures result in disastrous consequences for the organisation. An impact analysis of all the security candidates is therefore essential in the decision process to select appropriate countermeasures for a given attack.

In this paper, we propose a novel and systematic method to select security countermeasures from a pool of candidates, by ranking them based on the trade-off between their efficiency in stopping the attack, and their ability to preserve, at the same time, the best service to legitimate users. The method includes industrial evaluation and simulations of the impact associated to a given security measure.

The rest of the paper is structured as follows: Section \ref{sota} introduces the state of the art on service protection level, and return on response investment. Section \ref{cm_impact} presents the security and financial countermeasure impact analysis. Section \ref{cm_quantif} details the quantification of the proposed countermeasure impact model. Section \ref{use_case} gives an example of a case study. Finally, conclusions and perspectives for future work are presented in Section \ref{conclu}.

\section{State of the Art} \label{sota}

\subsection{Cyber Protection Level}\label{spl}

The cyber protection level is an evolution of the safety integrity level (SIL) \cite{Ingrey07, Mitchel10}, which is defined as a relative level of risk reduction provided by a safety function.  A SIL is determined based on a number of quantitative factors (using methods such as: risk matrices, risk graphs, layers of protection analysis) in combination with qualitative factors such as development process and safety life cycle management.

The cyber protection level refers to the strength of cyber security means deployed against a particular threat. The process is generally used to identify assets, threats, vulnerabilities, likelihood, countermeasures, and consequences. This is usually obtained from a risk analysis, following any of  the international standards (e.g., NIST \cite{NIST12}, ISO \cite{ISO08}), or any of the risk management methodologies (e.g., MEHARI \cite{mehari} and EBIOS \cite{ebios}) as well as expert knowledge. In our study, we consider the EBIOS methodology, defined by the French National Security Agency (ANSSI)\cite{ebios}.

The analysis follows several steps: 1) the context definition that determines stakeholders, processes, assets and dependencies, threat sources and existing security; 2) feared events and threat scenarios, with impact and occurrence probability; 3) risk evaluation that takes into account the existing security described in (1); 4) necessary measures related to risk mitigation.

Although most organizations follow a particular methodology to deploy a risk analysis, current approaches present several shortcomings: they rarely propose calculation methods for the protection level; none of them can be applied on an operational environment with ``living'' protection means (i.e., potentially unavailable for a period of time); they do not consider the different instances that must be deployed in the network to cover the threat everywhere; the effectiveness of a protection function is hardly considered in the analysis.

\subsection{Cost Sensitive Metrics}
Cost sensitive metrics are widely proposed as a viable approach to find an optimal balance between intrusion damages and response costs, and to guarantee the choice of the most appropriate response without sacrificing the system functionalities.

\subsubsection{Return On Investment (ROI):}
The simplest and most used approach for evaluating financial consequences of business investments, decisions and/or actions is the ROI metric. ROI is a relative measure that compares the benefits versus the costs obtained for a given investment \cite{Jeffrey04, Schmidt11}.

ROI basically shows how much a company earns from invested money. This metric supports decision makers in selecting the option(s) that have the highest return. ROI is calculated as the present value of accumulated net benefits over a certain time period minus the initial costs of investment, then divided by the initial costs of investment, as shown in Equation \ref{eq:roi}.

\begin{equation}\label{eq:roi}
ROI = \frac{B_t - C_t}{C_t} \times 100
\end{equation} \par

\noindent Where:

\noindent $B_t$ refers to all benefits during period t,\par
\noindent $C_t$ refers to all costs during period t\par

The decision rule is that the higher the ROI value, the more interesting the investment. However, Jeffery \cite{Jeffrey04} agrees that the major problem with ROI is that the metric does not include the time value of money, i.e., a 100\% ROI realized 1 year from today is more valuable than a 100\% ROI realized over 5 years. Furthermore, the costs and benefits of the project may vary over time, meaning that the cash flows are different in each time period. As a result, ROI is not a convenient way to compare projects when the costs and benefits vary with time, and it is also not useful for comparing projects that will run over different periods of time.

\subsubsection{Return On Security Investment (ROSI):}
It is a relative metric that compares the differences between the damages caused by attacks (before and after countermeasures) against the cost of the countermeasure \cite{Lockstep04, Sonner06, Kosutic}. To calculate ROSI, a formula adapted from the ROI metric is presented in Equation \ref{eq:rosi}.

\begin{equation}\label{eq:rosi}
ROSI = \frac{(ALE_b - ALE_a) - Cost_{CM}}{Cost_{CM}} \times 100
\end{equation} \par

\noindent Where:

\noindent $ALE_b$ refers to the annual loss expectancy before countermeasure,\par
\noindent $ALE_a$ refers to the annual loss expectancy after countermeasure,\par
\noindent $Cost_{CM}$ is the cost of the countermeasure\par

The calculation of each parameter composing the ROSI equation has been widely discussed by Lockstep Consulting \cite{Lockstep04}, and Kosutic \cite{Kosutic}. The former proposes a methodology that considers different levels of likelihood and severity, which are then, respectively transformed into frequency and direct cost; the latter considers on the one hand, parameters associated to the incident (e.g., financial losses, costs, frequency), and on the other hand, parameters associated to the protection (e.g., cost, benefits, life expectancy of the security measure).

Similar to the ROI metric, the decision rule states that the higher the ROSI value, the more interesting the investment.

\subsubsection{Return On Response Investment (RORI):}\label{rori}
It  is a service dependency index for cost sensitive response based on a financial comparison of the response alternatives \cite{Kheir10}. RORI is an adaptation of the ROSI index \cite{Sonner06} that provides a qualitative comparison of response candidates against an intrusion. RORI considers not only response effects on intrusions, but also response collateral damages as depicted in Equation \ref{eq:rori_Nizar}.

\begin{equation}\label{eq:rori_Nizar}
RORI = \frac{[IC_b - RC] - OC}{CD + OC} \times 100
\end{equation} \par

\noindent Where:

\noindent $IC_b$ represents intrusion impacts when no response is enforced,\par
\noindent $RC$ refers to the combined impact of intrusion and response,\par
\noindent $OC$ are operational costs that cover low level investments such as response setup and deployment
costs\par
\noindent $CD$ refers to collateral damages, which are costs that are added by a new response, and are not
related to intrusion costs\\

The deployment of the RORI index into real world scenarios has presented the following shortcomings:

\begin{itemize}
\item The RORI index is not defined when no countermeasure is selected. Since the operational cost (OC) is associated to the security measure, the RORI index will lead to an indeterminacy when no solution is enforced (NOOP).
\item The RORI index is not normalized with respect to the size and complexity of the infrastructure.
\item The absolute value of parameters such as $IC_b$ and $RC$ is difficult to estimate, whereas a ratio of these parameters is easier to determine, which in turn reduces errors of magnitude \cite{Gonzal14}.
\end{itemize}


\section{Countermeasure Impact Analysis} \label{cm_impact}

\subsection{Security Impact}\label{sec_impact}
Taking into account current shortcomings in risk analysis methodologies, we propose to evaluate a protection level against a threat, related to confidentiality, integrity, and availability, that considers technical ans business services as assets.

The protection level (PL) of a service $S_i$ against a threat $T_k$ is calculated using Equation \ref{eq:pl}.

\begin{equation}\label{eq:pl}
PL(S_i, T_k)=100-max (0, AD - AP)
\end{equation}

Where:
\begin{itemize}
\item AD = Assessed Danger, which represents the threat dangerousness in terms of confidentiality, integrity and availability (i.e., $d_{Ck}$, $d_{Ik}$, $d_{Ak}$), as well as the service value in terms of confidentiality, integrity and availability  (i.e., $v_{Ci}$, $v_{Ii}$, $v_{Ai}$), as shown in Equation \ref{eq:pl1}.

\begin{equation}\label{eq:pl1}
AD = \frac{(d_{Ck}\times v_{Ci})+(d_{Ik}\times v_{Ii})+(d_{Ak}\times v_{Ai})} {75}\times 100
\end{equation}

From Equation \ref{eq:pl1}, $d_{Ck}\times v_{Ci}$ represents the confidentiality-related impact, $d_{Ik}\times v_{Ii}$ represents the integrity-related impact, and $d_{Ak}\times v_{Ai}$ represents the availability-related impact. The more dangerous the threat and/or the more important the service in terms of confidentiality, integrity, and availability, the higher the impact. Values of dangerousness $d$  and service $v$ range from 0 to 5, therefore, the maximum possible value for AD is 75. We multiply this result by 100/75 in order to get homogeneous values of protection (assessed in a scale from 0 to 100).\\

The current proposal only considers the CIA services (i.e., confidentiality, integrity, and availability) for the calculation of the assessed danger, mainly due to two reasons: firstly, because the data is considered by the EBIOS methodology, and secondly, because the approach is used by our industrial partners at Cassidian Cybersecurity. However, it is also possible to use other parameters (e.g., criticality, accessibility, recuperability, vulnerability) as long as we are able to estimate their values for the selected threats and services.

\item AP = Assessed Protection, which represents the protection assigned against the threat k for the service i
(i.e., $p_{ik}$), as well as the effectiveness of the protection assigned against the threat k for the service i
(i.e., $e_{ik}$), as shown in Equation \ref{eq:pl2}.
\end{itemize}
\begin{equation}\label{eq:pl2}
AP = e_{ik}\times p_{ik}
\end{equation}

From Equation \ref{eq:pl2}, the assessed protection is estimated by experts. The effectiveness factor $e_{ik}$ is calculated depending on the type and distribution of protection (e.g., false positive rates, coverage of the network, feedback from operational teams, subjective figure).

To measure the impact of changes on security (SI, for Security Impact), we compare current and potential situations as depicted in Equations \ref{eq:si1} and \ref{eq:si2}

\begin{equation}\label{eq:si1}
SI = PL_{potential}(S_i, T_k) - PL_{current}(S_i, T_k)
\end{equation}

Where,
\begin{itemize}
\item $PL_{potential}$ is the protection level with a modified protection capacity against the threat $T_k$.
\end{itemize}

Equation \ref{eq:si2} provides more details about the delta between the current and the potential situation.
\begin{eqnarray}\label{eq:si2}
SI =& max \left(0, \left[ \frac{(d_{Ck}\times v_{Ci}) + (d_{Ik}\times v_{Ii}) + (d_{Ak}\times v_{Ai})} {75}\times 100 \right]- e_{ik}\times p_{ik} \right) - \nonumber \\
& max \left(0, \left[ \frac{(d_{Ck}\times v_{Ci}) + (d_{Ik}\times v_{Ii}) + (d_{Ak}\times v_{Ai})} {75}\times 100 \right]- e'_{ik}\times p'_{ik} \right) \nonumber \\
\end{eqnarray}

The variation is being on the protection $p_{ik}$ and/or its effectiveness $e_{ik}$.

\subsection{Financial Impact}\label{fin_impact}
An improvement of the RORI index has been proposed, taking into account not only the countermeasure cost and its associated risk mitigation, but also the infrastructure value and the expected losses that may occur as a consequence of an intrusion or attack \cite{Gonzal14}.

The improved RORI index handles the choice of applying no countermeasure to compare with the results obtained by the implementation of security solutions (individuals and/or combined countermeasures), and provides a response that is relative to the size of the infrastructure. The improved Return on Response Investment (RORI) index is calculated according to Equation \ref{eq:rori2}.

\begin{equation}\label{eq:rori2}
RORI = \frac{(ALE \times RM)-ARC}{ARC+AIV} \times 100
\end{equation}

Where:
\begin{itemize}
\item ALE is the Annual Loss Expectancy and refers to the impact cost obtained in the absence of security measures. ALE is expressed in currency per year (e.g., \$/year) and and includes loss of assets (La), loss of data (Ld), loss of reputation (Lr), legal procedures (Lp), loss of revenues from existing clients or customers (Lrec), loss of revenue from potential clients (Lrpc), other losses (Ol), contracted insurance (Ci), and the annual rate of occurrence (ARO), i.e., $ALE = (La+Ld+Lr+Lp+Lerc+Lrpc+Ol-Ci) \times ARO$\\

\item RM refers to the Risk Mitigation level associated to a particular solution. RM is defined from the security impact as a percentage (i.e., 0\% $\leq$ RM $\leq$ 100\%) that represents the additional countermeasure effectiveness over the best solution to be implemented in order to totally eradicate the threat. RM includes the protection level of potential and current situations, as presented in Section \ref{sec_impact} (i.e., RM = $\frac{PL_{potential}(S_i, T_k) - PL_{current}(S_i, T_k)}{100 - PL_{current}(S_i, T_k)} $)\\

\item ARC is the Annual Response Cost that is incurred by implementing a new security action. ARC = OC+CD from Equation \ref{eq:rori_Nizar}. ARC is always greater than or equal to zero ($ARC \geq 0$), and it is expressed in currency per year (e.g., \$/year). ARC includes Direct costs: e.g., Cost of implementation (Ci), Cost of maintenance (Cm), Other direct costs  (Odc); and Indirect costs (Ic), i.e., $ARC = Ci + Cm + Odc + Ic$\\

\item AIV is the Annual Infrastructure Value (e.g., cost of equipment, services for regular operations) that is expected for the system, regardless of the implemented countermeasures. AIV is greater than zero ($AIV > 0$), and it is expressed in currency per year (e.g., \$/year). AIV includes the following costs: Equipment Costs (Ec), Personnel costs (Pc), Service costs (Sc), Other costs (Oc), and Resell Value (Rv), i.e., $AIV = Ec + Pc + Sc + Oc - Rv$
\end{itemize}

\section{Countermeasure Impact Quantification} \label{cm_quantif}
The quantification of the parameters composing the RORI model proposed in Equation \ref{eq:rori2} is a task that requires expert knowledge, statistical data, simulation and risk assessment tools. Our experience in quantifying impact losses, as well as countermeasure costs and benefits for different security systems demonstrate that within 3 to 4 hours of discussions with use case providers and simple simulation runs, we are able to estimate each parameter composing the RORI model. The remaining of this section proposes a simple and well structured methodology to help security analysts in the estimation of such parameters.

\subsection{Annual Loss Expectancy}\label{ale}

For the estimation of the ALE, we adopted the approach proposed by Lockstep \cite{Lockstep04} to use the severity scale of values, which convert qualitative estimations into quantitative values of costs. For instance, a `minor' loss of assets (La) represents a cost of \$1,000; whereas a `serious' loss of assets (La) represents a cost of \$1,000,000. The estimation of all other losses (i.e., Ld, Lr, Lp, Lrec, Lrpc, Ol) follows the same approach.

The likelihood of an incident is transformed into a frequency value, which results into the Annual Rate of Occurrence (ARO) parameter. For instance, a `low likelihood' means that the incident is likely to occur once every year, (ARO = 1); whereas, a `high likelihood' means that the incident is likely to occur once per month or less, (ARO = 12).

Both parameters (i.e., severity and likelihood) are estimated using a survey and scoring system, which combine expert knowledge and statistical data to quantify risk exposure. In order to handle uncertainty, we use Monte Carlo simulation. To run our simulation, we chose triangular distributions to evaluate the most likely values assigned to each level of security and likelihood, with minimum and maximum possible values of each level. This type of statistical computations can be easily achieved using basic statistical software or spreadsheet editors\footnote{Quadrant: The Quick and dirty risk analysis tool, available at: www.qdrnt.com/home.htm}\footnote{Monte Carlo simulation for excel featuring distribution strings, available at: http://xlsim.com/xlsim/index.html}.

After 250 iterations, we were able to obtain a value of the losses and frequency that compose the ALE parameter, which represents the expected annual loss as a consequence of the realization of a given threat.

\subsection{Risk Mitigation}\label{rm}

A risk analysis, as performed by Cassidian cyber-security experts\footnote{http://www.cassidiancybersecurity.com/en\_US/web/guest/cybersecurity}, gives the list of threats directly endangering business and technical services of the entity to protect, and the available protection means. The level of protection related to a given set of services is assessed using different kinds of information:
\begin{enumerate}
\item Types of security devices able to detect and/or react against an activity related to a threat occurrence; given by cyber-security experts.
\item  Instances of security devices actually deployed to protect services; given by security architects
\end{enumerate}

Services are modelled using dependency models. Identified threats and related protection measures (if they exist) are associated to services. We obtain, for each service a list of couples (threat, protection).

A threat is characterized by a dangerousness level in terms of confidentiality (i.e., $d_{Ck}$), integrity (i.e., $d_{Ik}$) and availability (i.e., $d_{Ak}$). We consider each service and their value as per confidentiality (i.e., $v_{Ci}$), integrity (i.e., $v_{Ii}$) and availability (i.e., $v_{Ai}$) in order to determine the potential effect of threats on services. An example of the asset values is represented in Table \ref{table:ttmatrix}.

\begin{table}
\begin{center}
\caption{Assessed Dangerousness Matrix}
  \label{table:ttmatrix}
\begin{tabular}{|m{0,5cm}|m{0,5cm}|m{0,5cm}|c |c|c|c|c|c|}
 \cline{5-9}
 \multicolumn{4}{r|}{C}&5&5&0&4&3\\
 \cline{5-9}
 \multicolumn{4}{r|}{I}&5&5&0&4&3\\
 \cline{5-9}
 \multicolumn{4}{r|}{A}&5&5&2&4&3\\
 \cline{5-9}
\multicolumn{3}{r}{}&&&&&&\\
\multicolumn{3}{r}{}&&&&&&\\
\multicolumn{3}{r}{}&&&&&&\\
\multicolumn{1}{c}\bf{C} &  \multicolumn{1}{c}{I}& \multicolumn{1}{c}{A} &&\rotatebox{90}{\parbox{2mm}{\multirow{3}{*}{Service1}}}&\rotatebox{90}{\parbox{2mm}{\multirow{3}{*}{Service2}}} & \rotatebox{90}{\parbox{2mm}{\multirow{3}{*}{Service3}}}& \rotatebox{90}{\parbox{2mm}{\multirow{3}{*}{Service4}}}& \rotatebox{90}{\parbox{2mm}{\multirow{3}{*}{Service5}}} \\
\hline
1&2&3&Threat1&40&40&8&32&24\\ \hline
3&3&3&Threat2&60&60&N/A&N/A&N/A\\ \hline
2&2&2&Threat3&N/A&N/A&5&32&24\\ \hline
5&5&5&Threat4&N/A&100&N/A&N/A&N/A\\ \hline
4&4&4&Threat5&N/A&N/A&N/A&N/A&48\\ \hline
5&5&5&Threat6&100&100&13&80&60\\ \hline
3&3&3&Threat7&60&60&8&36&36\\ \hline
2&2&0&Threat8&N/A&27&0&N/A&N/A\\ \hline
4&5&3&Threat9&80&N/A&N/A&N/A&N/A\\ \hline
3&3&3&Threat10&60&60&8&48&36\\
\hline
\end{tabular}
\end{center}
\end{table}


Dangerousness levels and values are integers ranging from 0 (meaning respectively no danger / no value) to 5 (meaning respectively highest danger / biggest value). Dangerousness and asset values are given by experts.
Cell values are calculated using the AD such as described in Equation \ref{eq:pl1}. Highest threat effect would be 75 (dot product of danger level and service value per CIA criteria). The result is finally reported as a percentage. It is important to note that the ``N/A'' flag in some cells means the threat does not endanger the service.


A protection ($p_{ik}$) is characterized by its effectiveness ($e_{ik}$) to prevent a threat from occurring. This is an integer ranging from 0 to 100. The protection either exists ($p_{ik} = 1$) or does not exist ($p_{ik} = 0$). When the protection is different from 0, the related threats are supposed to be mitigated by some of the protection means described in the service model.

Table \ref{table:ptmatrix} depicts an example of protection capacity on different services affected by several threats. We identify the services at which protection measures have been deployed, and their ability to mitigate threats. As a result, we consider the actual danger being the difference of threat level and protection level. Results of the aforementioned example are depicted in Table \ref{table:admatrix}.\\

\begin{table}[!htb]
    \begin{minipage}{.4\linewidth}
      \caption{Protection Capacity} \label{table:ptmatrix}
      \centering
\begin{tabular}{|c|m{0,5cm}|m{0,5cm}|m{0,5cm}|m{0,5cm}|m{0,5cm}|}
\cline{2-6}
\multicolumn{1}{c|}{}&&&&&\\
\multicolumn{1}{c|}{}&&&&&\\
\multicolumn{1}{c|}{}&&&&&\\
\multicolumn{1}{c|}{}&\rotatebox{90}{\parbox{2mm}{\multirow{3}{*}{Service1}}}&\rotatebox{90}{\parbox{2mm}{\multirow{3}{*}{Service2}}} & \rotatebox{90}{\parbox{2mm}{\multirow{3}{*}{Service3}}}& \rotatebox{90}{\parbox{2mm}{\multirow{3}{*}{Service4}}}& \rotatebox{90}{\parbox{2mm}{\multirow{3}{*}{Service5}}} \\
\hline
Threat1&&&&&\\ \hline
Threat2&75&75&&&\\ \hline
Threat3&&&60&60&60\\ \hline
Threat4&&&&&\\ \hline
Threat5&&&&&40\\ \hline
Threat6&100&100&100&100&100\\ \hline
Threat7&50&&50&&50\\ \hline
Threat8&&&&&\\ \hline
Threat9&&&&&\\ \hline
Threat10&90&90&90&90&90\\
\hline
\end{tabular}

\end{minipage}%
    \begin{minipage}{.7\linewidth}
      \centering
        \caption{Actual Danger Matrix}\label{table:admatrix}

\begin{tabular}{|c|c|c|c |c|c|c|c|c|m{0,5cm} l}
 \cline{5-9}
 \multicolumn{4}{r|}{C}&5&5&0&4&3&\\
 \cline{5-9}
 \multicolumn{4}{r|}{I}&5&5&0&4&3&\\
 \cline{5-9}
 \multicolumn{4}{r|}{A}&5&5&2&4&3&\\
 \cline{5-9}
\multicolumn{3}{r}{}&&&&&&&\\
\multicolumn{3}{r}{}&&&&&&&\\
\multicolumn{3}{r}{}&&&&&&&\\
\multicolumn{1}{c}\bf{C} &  \multicolumn{1}{c}{I}& \multicolumn{1}{c}{A} &&\rotatebox{90}{\parbox{2mm}{\multirow{3}{*}{Service1}}}&\rotatebox{90}{\parbox{2mm}{\multirow{3}{*}{Service2}}} & \rotatebox{90}{\parbox{2mm}{\multirow{3}{*}{Service3}}}& \rotatebox{90}{\parbox{2mm}{\multirow{3}{*}{Service4}}}& \rotatebox{90}{\parbox{2mm}{\multirow{3}{*}{Service5}}} &&\\
\cline{1-9}
1&2&3&Threat1&40&40&8&32&24&&\\ \cline{1-9}
3&3&3&Threat2&-15&-15&N/A&N/A&N/A&&\\ \cline{1-9}
2&2&2&Threat3&N/A&N/A&-55&-28&-36&&\\ \cline{1-9}
5&5&5&Threat4&N/A&100&N/A&N/A&N/A&&\\ \cline{1-9}
4&4&4&Threat5&N/A&N/A&N/A&N/A&8&&\\ \cline{1-9}
5&5&5&Threat6&0&0&-87&-20&-40&&\\ \cline{1-9}
3&3&3&Threat7&10&60&-42&36&-14&&\\ \cline{1-9}
2&2&0&Threat8&N/A&27&0&N/A&N/A&&\\ \cline{1-9}
4&5&3&Threat9&80&N/A&N/A&N/A&N/A&&\\ \cline{1-9}
3&3&3&Threat10&-30&-30&-82&-42&-54&&\\
\cline{1-9}
\end{tabular}
\end{minipage}
\end{table}

Empty cells from Table \ref{table:ptmatrix} mean that protection does not exist in such service against a particular threat. Cells from Table \ref{table:admatrix} show the actual danger. In order to obtain the PL value in each cell, we use Equation \ref{eq:pl}

\subsection{Annual Response Cost}\label{arc}
In contrast to the AIV parameter, the ARC is a variable cost associated with the implementation of a given countermeasure. For instance, let us suppose that the user authentication information of a Web service is stored in a database. Whenever users want to access the system, they need to provide their corresponding login and password. However, for suspicious users, the organization wants to implement a countermeasure that asks for a double authentication (e.g., a challenge question, a security pin). The implementation of this countermeasure requires the organization to spend additional employee-hours which in turn represents a given cost. This latter is defined as the cost of implementation (Ci).

In addition, the countermeasure is going to be active only for suspicious users for a given period of time, which means that the system will turn the countermeasure from `on' to `off' according to the security tests and analysis performed. These tests and analysis represent the cost of maintenance (Cm) to the organization.

The activation/deactivation of a given countermeasure engenders other direct and indirect costs. For instance, requesting an additional authentication method to legitimate users may cause these users to unsubscribe from the service and search for another one. This collateral damage represents an indirect cost (Ic) to the organization. Collateral damages can be quantified as the variation between the current and the projected productivity that an organization experiences due to a side effect of a given solution \cite{Sonner06}.

\subsection{Annual Infrastructure Value}\label{aiv}

This parameter is calculated as the sum of the annual value of all the equipments, i.e., Policy Enforcement Points (PEP), that are needed to be deployed in the preliminary phase of the system architecture in order to guarantee a desired level of security. The AIV includes the cost of purchasing, licensing, and/or leasing the security equipments in a given organization.

It is important, however, to answer the following questions while estimating the AIV parameter:
\begin{itemize}
\item What kind of PEPs (e.g., Firewalls, IPS, IDS, SIEM) and which quantity is required for the system security?
\item What is the lifetime expectancy of the PEP?
\item What is the PEP's deployment time?
\item What is the annual cost of purchase, licensing or leasing of the PEP?
\item How many employee-hours are required for the operation of the PEP?
\item How long (i.e., hours/year) is the PEP expected to be active?
\item Is there an insurance contracted for the PEP? If so, how much does it cost per year?
\item How frequently (i.e., times/year) does the PEP need to be checked or maintained?
\item Is there any other cost associated with the operation of the PEP in the security infrastructure?
\item What is the amortization value of the PEP?
\end{itemize}

\section{Use Case} \label{use_case}

This section describes a simple case study provided by Cassidian CyberSecurity, the cyber security company of the Airbus group, and a major provider of global security solutions and services.

The scenario is based on a risk analysis performed on a company, limited to three services. The security auditors have determined the value of these services for the company, taking into account the company activity, stakeholders, technical and human constraints (e.g., skill level of the personnel in terms of security-related good practices), the loss of money in case of failures, etc. The risk analysis has been performed according to the EBIOS methodology. Four threats  have been considered in this study. Their effects on targeted elements enable the auditors to evaluate the dangerousness criteria. Countermeasures have been proposed by the auditors to make the risk level acceptable along the company criteria.

The subsequent deployment of security devices compliant with the experts recommendations leads to provide the following matrices : threat target matrix (Table \ref{table:ttmatrix1}), and protection capacity matrix (Table \ref{table:ptmatrix1}).

\begin{table}[!htb]
    \begin{minipage}{.5\linewidth}
      \caption{Threat Target Matrix} \label{table:ttmatrix1}
      \centering

\begin{tabular}{|m{0,5cm}|m{0,5cm}|m{0,5cm}|m{2cm} |m{0,75cm}|m{0,75cm}|m{0,75cm}|}
 \cline{5-7}
 \multicolumn{4}{r|}{C}&0&5&5\\
 \cline{5-7}
 \multicolumn{4}{r|}{I}&5&5&5\\
 \cline{5-7}
 \multicolumn{4}{r|}{A}&5&5&3\\
 \cline{5-7}
\multicolumn{3}{r}{}&&&&\\
\multicolumn{3}{r}{}&&&&\\
\multicolumn{3}{r}{}&&&&\\
\multicolumn{3}{r}{}&&&&\\
\multicolumn{3}{r}{}&&&&\\
\multicolumn{3}{r}{}&&&&\\
\multicolumn{3}{r}{}&&&&\\
\multicolumn{3}{r}{}&&&&\\
\multicolumn{1}{c}\bf{C} &  \multicolumn{1}{c}{I}& \multicolumn{1}{c}{A} &&\rotatebox{90}{\parbox{2mm}{\multirow{3}{*}{Web services}}}&\rotatebox{90}{\parbox{2mm}{\multirow{3}{*}{Network infrastructure}}} & \rotatebox{90}{\parbox{2mm}{\multirow{3}{*}{User service}}}\\
\cline{1-7}
1&3&2&Web site sabotage&33&N/A&N/A\\ \cline{1-7}
3&1&5&Network infrastructure attack&N/A&60&N/A\\ \cline{1-7}
5&4&2&User workstation compromise&N/A&N/A&68\\ \cline{1-7}
5&4&3&Admin workstation compromise&N/A&N/A&72\\ \cline{1-7}
\end{tabular}
\end{minipage}%
    \begin{minipage}{.6\linewidth}
      \centering
        \caption{Protection Capacity Matrix}\label{table:ptmatrix1}

\begin{tabular}{|m{2cm}|m{0,75cm}|m{0,75cm}|m{0,75cm}|}
\cline{2-4}
\multicolumn{1}{c|}{}&&&\\
\multicolumn{1}{c|}{}&&&\\
\multicolumn{1}{c|}{}&&&\\
\multicolumn{1}{c|}{}&&&\\
\multicolumn{1}{c|}{}&&&\\
\multicolumn{1}{c|}{}&&&\\
\multicolumn{1}{c|}{}&&&\\
\multicolumn{1}{c|}{}&&&\\
\multicolumn{1}{c|}{}&\rotatebox{90}{\parbox{2mm}{\multirow{3}{*}{Web services}}}&\rotatebox{90}{\parbox{2mm}{\multirow{3}{*}{Network infrastructure}}} & \rotatebox{90}{\parbox{2mm}{\multirow{3}{*}{User service}}} \\
\hline
Web site sabotage&50&&\\ \hline
Network infrastructure attack&&80&\\ \hline
User workstation compromise&&&17\\ \hline
Admin workstation compromise&&&50\\ \hline

\end{tabular}
\end{minipage}
\end{table}
The main danger on user and admin workstations lies in their compromission by malware programs. To counter this threat we deploy a protection with an effectiveness assessed by experts e = 50\%. The effectiveness value is obtained considering several criteria:
\begin{itemize}
\item reliability of the malware detection software: the cyber company leading audits maintain a knowledge base regarding the reliability of security products. Particularly, anti-virus system reliability has been tested against malwares discovered and published within a period of 6 months. These tests are possible using online services such as VirusTotal. With an up-to-date base, 80\% of the injected malware programs were detected by the malware detection tool deployed in the audited company. Then reliability score is 80\%.
\item signature base update policy: the frequency is set to one per week, which is assessed being far from achieving complete protection, therefore the score is set to 60\%. The following scale is being used: 100\% daily update, 60 \% weekly update, 20\% monthly update, 5\% annual update, 1\% no update since installation.
\item resilience : this one is 100\%, as tests over a 1 month period do not reveal any dysfunction. Indeed the cyber security company periodically launches test campaigns on security product resilience.
\end{itemize}
The effectiveness is then evaluated as the product of the reliability, policy and resilience scores (i.e., e = reliability\_score $\times$ update\_policy\_score $\times$ resilience\_score). This gives a result of 48\%, approximated to 50\%. This kind of protection is deployed on every administration workstation, and in only 1/3 user workstations (900 PC among 2700 for the whole organization), mainly for cost reasons.
The protection level (PL) is calculated using Equation \ref{eq:pl}, as follows:

PL (User service, User workstation compromise) = 100 - (68 - 17)

PL (User service, User workstation compromise) = 100 - 51 = 49

A malware is detected on a user computer (among those unprotected). The proposed countermeasure consists on deploying an anti-malware agent on it and extend the solution to the other 1,800 workstations. The technical assessment of the countermeasure is shown in Tables \ref{table:fpcmatrix} and \ref{table:pdmatrix1}.

	\begin{table}[!htb]
    \begin{minipage}{.3\linewidth}
      \caption{Foreseen Protection Capacity Matrix} \label{table:fpcmatrix}
      \centering
\begin{tabular}{|m{2cm}|m{0,75cm}|m{0,75cm}|m{0,75cm}|}
\cline{2-4}
\multicolumn{1}{c|}{}&&&\\
\multicolumn{1}{c|}{}&&&\\
\multicolumn{1}{c|}{}&&&\\
\multicolumn{1}{c|}{}&&&\\
\multicolumn{1}{c|}{}&&&\\
\multicolumn{1}{c|}{}&&&\\
\multicolumn{1}{c|}{}&&&\\
\multicolumn{1}{c|}{}&&&\\
\multicolumn{1}{c|}{}&\rotatebox{90}{\parbox{2mm}{\multirow{3}{*}{Web services}}}&\rotatebox{90}{\parbox{2mm}{\multirow{3}{*}{Network infrastructure}}} & \rotatebox{90}{\parbox{2mm}{\multirow{3}{*}{User service}}} \\
\hline
Web site sabotage&50&&\\ \hline
Network infrastructure attack&&80&\\ \hline
User workstation compromise&&&50\\ \hline
Admin workstation compromise&&&50\\ \hline

\end{tabular}
\end{minipage}%
    \begin{minipage}{.8\linewidth}
      \centering
        \caption{Potential Danger Matrix}\label{table:pdmatrix1}

\begin{tabular}{|m{0,5cm}|m{0,5cm}|m{0,5cm}|m{2cm}|m{0,75cm}|m{0,75cm}|m{0,75cm}|}
 \cline{5-7}
 \multicolumn{4}{r|}{C}&0&5&5\\
 \cline{5-7}
 \multicolumn{4}{r|}{I}&5&5&5\\
 \cline{5-7}
 \multicolumn{4}{r|}{A}&5&5&3\\
 \cline{5-7}
\multicolumn{3}{r}{}&&&&\\
\multicolumn{3}{r}{}&&&&\\
\multicolumn{3}{r}{}&&&&\\
\multicolumn{3}{r}{}&&&&\\
\multicolumn{3}{r}{}&&&&\\
\multicolumn{3}{r}{}&&&&\\
\multicolumn{3}{r}{}&&&&\\
\multicolumn{3}{r}{}&&&&\\
\multicolumn{1}{c}\bf{C} &  \multicolumn{1}{c}{I}& \multicolumn{1}{c}{A} &&\rotatebox{90}{\parbox{2mm}{\multirow{3}{*}{Web services}}}&\rotatebox{90}{\parbox{2mm}{\multirow{3}{*}{Network infrastructure}}} & \rotatebox{90}{\parbox{2mm}{\multirow{3}{*}{User service}}}\\
\cline{1-7}
1&3&2&Web site sabotage&-17&N/A&N/A\\ \cline{1-7}
3&1&5&Network infrastructure attack&N/A&-20&N/A\\ \cline{1-7}
5&4&2&User workstation compromise&N/A&N/A&18\\ \cline{1-7}
5&4&3&Admin workstation compromise&N/A&N/A&22\\ \cline{1-7}
\end{tabular}
\end{minipage}
\end{table}
The risk mitigation RM (user workstation compromise) = (82-49) / 51 = 65\%. The anti-malware editor cost policy is the following: 40,000\euro \hspace{1mm} per year for a maximum of 2,000 embedded agents; 50,000\euro \hspace{1mm} per year for a maximum of 5,000 embedded agents.

Considering that the annualized loss expectancy for a malware in the system is estimated at ALE = 100,000\euro \hspace{1mm} per year, and that the annual infrastructure value is estimated at AIV = 75,000\euro \hspace{1mm} per year, we use Equation \ref{eq:rori2} to perform the countermeasure evaluation. Table \ref{table:SingleResults1} shows these results, and details information regarding countermeasure cost, mitigation level, and RORI index.

From Table \ref{table:SingleResults1}, the first candidate (i.e., C1) proposes to accept the risk by doing no operation (NOOP). This alternative does not provide any mitigation level (RM=0) and does not generate any additional cost (ARC=0). The expected return on the response investment is therefore null (RORI=0).

The second alternative (i.e., C2) proposes to install agents only in the infected host. This alternative will not change the danger of the total group of 2,700 hosts. The mitigation level will be therefore close to zero (RM=0.01). However, taking into account that a license to install an anti-malware agent for a maximum of 2,000 hosts is already being paid, the ARC value to be installed in 1 additional host will only consider the cost of deployment (e.g., deploying a license in one host takes in average 10 minutes, and 1 employee-hour costs 100\euro \hspace{1mm} at Cassidian Cyber Security), therefore ARC(1 host) = 17\euro.

The third alternative (i.e., C3) suggests to install agents in 1,100 additional workstations (the maximum number of hosts allowed by the license). The mitigation level is calculated considering the current protection level (\textit{PLcurrent} = 49), and the potential protection level (\textit{PLpotentiel} = 100- max (0, 68-50*2000/2700) = 69), therefore RM = (69 -49) / (100 -49) = 39\%. In addition, the ARC for 1,100 additional hosts (to reach the 2,000 agents limit) is equivalent to ARC= 1100 x 17 = 18700 \euro . As a result, the return on response investment is equivalent to (RORI = 21.66\%).

The fourth evaluated candidate proposes to install agents in every administration workstation of the whole organization (i.e., 2,700 workstations). This requires to pay an additional of 10,000\euro, for a license that will allow to install agents in a maximum of 5,000 hosts. The mitigation level is calculated considering the current protection level (\textit{PLcurrent} = 49), and the potential protection level (\textit{PLpotentiel} = 100- max (0, 68-50*2700/2700) = 82), therefore RM = (82 -49) / (100 -49) = 65\%. In addition, the ARC for 1,800 additional hosts (to reach the 2,700 agents) is equivalent to ARC= 1800 x 17 = 30,600\euro + 10,000\euro = 40,600\euro. As a result, the return on response investment is equivalent to (RORI = 21.11\%).

\begin{table}
\begin{center}

\caption{Countermeasure Evaluation against Malware Infection}
  \label{table:SingleResults1}
\begin{threeparttable}

\begin{tabular}{p{7cm}p{2cm}p{1cm}p{1cm}}
\hline
\hline
\bf{Countermeasure} &  \bf{ARC} & \bf{RM} &\bf{RORI}\\
\hline
C1. No operation (NOOP) &  0 \euro &0.00 &0.00\% \\
C2. Install agent only in the infected hosts &  17 \euro &0.01 &1.31\% \\
C3. Install agents in 1,100 hosts (to reach the 2000 agent-limit) &  18,700 \euro &0.39 &21.66\% \\
C4. Install agents in 1,800 hosts (to protect all workstations) &  40,600 \euro &0.65 &21.11\% \\

\hline
\hline
\end{tabular}
\end{threeparttable}

\end{center}
\end{table}

After the evaluation of the different candidates to mitigate a malware
infection, we select alternative 3 as the optimal countermeasure,
since it provides the highest RORI index. C3 proposes to install
anti-malware agents in 1,100 hosts, additional to the already 900
protected hosts.

\section{Conclusions and Future Work}
\label{conclu}

We have proposed in this paper a novel and well structured method to
select security countermeasures from a pool of candidates, based on
their technical and financial impact. The method includes industrial
evaluation and simulations of the impact associated to a given
security measure.

By calculating the potential new protection level, we are able to
compare the current versus the potential change. As a result, we
obtain quantitative information on the improvement or degradation of
security at the service level. However, nowadays this function is
limited to the protection level measurement after the addition or
removal of protection measures in the network (e.g.,
enabling/disabling a security function will be considered as an
addition/removal security function). We do not support detailed
settings of security devices such as filtering rules in a firewall.

Future work will define the full service protection level as the
overall protection of services for the entity due to several reasons:
1) to be aware of the general security level; 2) because actions to
improve security for a service may have negative consequences to
others (e.g., move of a security device), or may decrease the
protection against other threats (e.g., replacement of a security
device).

Another task will consist in proposing guidelines for the protection
effectiveness per type of security function. This parameter is very
important in the proposed approach. Giving subjective value would ruin
the effort to rationalize the RORI result.

\subsubsection*{\bf{Acknowledgements:}} The research in this paper has
received funding from the Information Technology for European
Advancements (ITEA2) within the context of the ADAX Project (Attack
Detection and Countermeasure Simulation)

\bibliographystyle{eptcs}

\end{document}